\documentclass[structabstract]{aa}  
\usepackage{graphicx}
\usepackage{txfonts}
\begin{document}
    \title{$\tau$ Boo b: Hunting for reflected starlight\thanks{Based on observations made with ESO Telescopes 
   at the Paranal Observatory under programme ID 079.C-0413(A).}}


   \author{F. Rodler\inst{1,2,3}
          \and
          M. K\"urster\inst{4}
          \and
          T. Henning\inst{4}
          }
 
   \offprints{frodler@iac.es}
  \institute{Instituto de Astrof\'{i}sica de Canarias,
              C/V\'{i}a L\'{a}ctea s/n, 38205 La Laguna, Spain
     \and
   			formerly at the Max-Planck-Institut f\"ur Astronomie, 
              K\"onigstuhl 17, 69117 Heidelberg, Germany
         \and
             formerly at the Institut f\"ur Astronomie, Universit\"at Wien,
              T\"urkenschanzstrasse 17, A-1180 Vienna, Austria    
\and
  			Max-Planck-Institut f\"ur Astronomie, 
              K\"onigstuhl 17, 69117 Heidelberg, Germany}

  \date{Received ?; accepted ?}
 
 
  \abstract
   {}
   {We attempt to detect starlight reflected from the hot Jupiter orbiting the
   main-sequence star $\tau$~Boo, in order to determine the albedo of the
   planetary atmosphere, the orbital inclination of the planetary system and the
   exact mass of the planetary companion. }
   {We analyze high-precision, high-resolution spectra, collected over two half nights using UVES at the VLT/UT2, 
   by way of data synthesis. We interpret our data using two different atmospheric models for hot Jupiters.}
   {Although a weak candidate signal appears near the most probable radial velocity
     amplitude, its statistical significance is insufficient for us to claim a detection. 
     However, this feature agrees very well with a completely independently obtained result by another research group, which searched
     for reflected light from $\tau$~Boo~b.
     As a consequence of the non-detection of reflected light, we place upper limits to the planet-to-star flux ratio at the
   99.9\% significance level. For the most probable orbital inclination around $i=46^\circ$, 
    we can limit the relative reflected
     radiation to be less than $\epsilon =
   5.7\times10^{-5}$ for grey albedo.
       This
     implies a geometric albedo smaller than $0.40$, assuming a planetary radius of
     $1.2~\rm{R_{Jup}}$.  }
   {}

 
   \maketitle
%

\section{Introduction}

Since the detection of the first exoplanet orbiting a solar-type
star, more than 400 exoplanets have been detected.
 The existence of most 
of these planets was established by monitoring radial velocity (RV) variations of the host star, originating from the gravitational pull of the unseen planetary companion.
 So-called hot
Jupiters are giant planets only a few solar radii away from their
host stars that provide the opportunity to attempt the detection of starlight reflected
from these planets. Five extended campaigns for the search for
refected light by using high-resolution spectroscopy were completed by different groups
(Charbonneau et al. 1999, Collier Cameron et al. 1999, Collier Cameron et al. 2002, Leigh et al. 2003a., Leigh et al. 2003b, Rodler et al. 2008). 
Apart from Collier Cameron et al. (1999), who claimed a detection of reflected starlight, which was later withdrawn (Collier Cameron et al.
2000), all campaigns resulted in a non-detection of reflected starlight, and upper limits to the planet-to-star flux ratio 
and to the geometric albedo of these planets were established: to date, the tightest 99.9\% confidence upper limits on the
geoemtric albedos of the hot Jupiters $\tau$~Boo~b, HD~75289b and $\upsilon$~And~b 
 are $0.39$ (Leigh et al. 2003a), $0.46$ (Rodler et al. 2008), and $0.42$ (Collier Cameron et al. 2002), respectively.
These results provided 
important constraints on models of the planetary atmospheres such as those by Sudarsky et al. (2000, 2003). As a result,
 models that
predicted a high reflectivity for the planetary atmosphere could
be ruled out for some of the studied planets.

 More recently, the albedos of several transiting hot Jupiters at visual wavelengths could be further constrained from measurements of the secondary transit
events. For the transiting hot Jupiter HD~209458b, Rowe et al.~(2008) placed a very stringent $3\sigma$ upper limit 
on the geometric albedo of 0.17 in the wavelength regime of 400-700 nm. Using data of the CoRot-satellite of the transiting hot Jupiter CoRot-1b, 
Snellen et al.~(2009) measured
a phase-dependency of the planetary flux and reported an upper limit on the geometric albedo
of 0.2 in the wavelength range of 400-1000 nm. This upper limit of the geometric albedo of CoRot-1b was confirmed by an 
independent analysis by Alonso et al.~(2009a). 
Using the same data set for the same target, Rogers et al.~(2009) reports an estimate 
of the planetary geometric albedo to be 0.05. 
Alonso et al.~(2009b) placed a very stringent upper limit on the geometric albedo of 0.06 
of the hot Jupiter CoRot-2b in the wavelength range 400-1000 nm. The analysis of measurements of the secondary eclipse of 
the hot Jupiter HAT-P-7b with the EPOXI spacecraft and the 
Kepler-satellite led to an estimate of the geoemtric albedo of 0.13 at 650 nm (Christiansen et al.~2009). Measurements of the secondary transit of the planet Ogle-Tr-56 in the
$z'$-band indicate a low geometric albedo less than 0.15 (Sing \& Morales~2009). For a general summary of the secondary transit measurements
we refer to Cowan \& Agol (2010).

$\tau$ Bo\"otis (HD~120136A) is a main-sequence star of spectral type F7, located
at a distance of 15.6 pc from our Solar System. Butler et al.~(1997) detected a massive hot Jupiter orbiting
$\tau$~Boo via RV measurements; we note that this star is one of the brightest
stars in the sky harboring a planet. Shortly after the discovery of the planetary companion,
 two research groups started campaigns for the search for
reflected light by way of high-precision spectroscopy, which finally resulted
in
non-detections. The tightest published $99.9\%$ confidence upper limit
to the geometrical albedo is $p<0.39$ under the assumption that the planetary
radius $R_{\rm p}=1.2~R_{\rm Jup}$, orbital inclination values $i \geq 36^{\circ}$ and grey albedo
(Leigh et al.~2003a). These authors report the detection of a
candidate signal of marginal significance with a projected orbital RV
semi-amplitude $K_{\rm p}=97~{\rm km~s^{-1}}$, which corresponds to an orbital inclination
$i\approx 40^{\circ}$. 

Here, we present our analysis of new observations of the
   planetary system of $\tau$~Boo taken with the UV-Visual
   Echelle Spectrograph (UVES) mounted 
   on the VLT/UT2 at Cerro Paranal in Chile. 
Section~2 describes the
   basic ideas of the search for reflected light. 
   Section~3 provides an overview over the planetary system, while Section~4 outlines the acquisition and
   reduction of the high-resolution spectra. Section~5 provides a
   detailed description of the data analysis with a data modeling
   approach. Finally, in Section~6 we present the results, which are discussed in Section~7.

\section{Starlight reflected from the planet}

 High resolution spectroscopic searches 
utilize the fact that the spectrum reflected from the planet is essentially a copy of the
rich stellar absorption line spectrum except for the following differences:
\begin{itemize}

\item It is scaled down in intensity 
by more than five orders of magnitude in the visual for hot Jupiters (Section~\ref{A2.1}).
\item It is shifted in wavelength according to the relative orbital radial velocity of the planet
(Section~\ref{dop2}).

\item It displays a different degree of rotational broadening corresponding to the 
rotational velocity of the star as
seen from the hot Jupiter whose own rotation typically contributes very little to the line broadening (Section~\ref{2:rotbro})
\end{itemize}

 \subsection{Photometric Variations} \label{A2.1}
  For exoplanets, the enormous brightness contrast between the star and 
  the planet constitutes a considerable challenge when attempting to observe some 
  kind of direct signal from the planet.
  For close-in planets, such as hot
  Jupiters, the main contribution to the flux in the visual consists of the
  reflected starlight and not the intrinsic luminosity (Seager \& Sasselov 1998).
  
   According to Charbonneau et al. (1999), the amount of starlight
  reflected from a planet which is fully illuminated can be described by
\begin{equation} \label{equ:2}
     \epsilon(\lambda) = p(\lambda) \left(\frac{R_{\rm{p}}}{a}\right)^2 \rm{,}
  \end{equation}
  where $p(\lambda)$ denotes the albedo of the planet as a function of the
  wavelength $\lambda$, $R_{\rm{p}}$ the planetary radius and $a$ the
  star-planet separation.
The planetary
  radius of $\tau$~Boo~b is unknown; it can, however, be estimated from a comparison with the
  transiting planets, which provide an exact
  determination of their masses and radii (Bakos et al. 2007). The orbital
  radius is tightly constrained through Kepler's third law.
 

  As the planet orbits its star, the fraction of the illuminated
  part of the planet changes relatively to the observer. Consequently, the
  observed reflected light  is reduced, depending on the model describing the
  scattering behaviour of the atmosphere, its orbital inclination
  $i \in [0^{\circ},90^{\circ}]$ and the orbital phase $\phi \in [0, 1]$ of the planet. 
  We note that we adopt the convention that $\phi=0$ represents inferior
  conjunction  of the planet (for $i=90^{\circ}$, it would be the transit position). 
  Lacking observational data for hot Jupiters we apply an empirical
  scattering model of the atmospheres of Jupiter and Venus (Hilton 1992),
  which can be approximated by

  \begin{equation} \label{equ:3}
    \mu(\phi, i)= 10^{-0.4 \zeta(\alpha)}  \rm{,}
  \end{equation}
  where
  \begin{equation} \label{equ:5}
    \zeta(\alpha) = 0.09 \left(\frac{1.8~\alpha}{\pi}\right) +  2.39
    \left(\frac{1.8~\alpha}{\pi}\right)^2 -0.65
    \left(\frac{1.8~\alpha}{\pi}\right)^3  
  \end{equation}
  and the phase angle $\alpha$ is given by
  \begin{equation} \label{equ:1}
    \cos \alpha = -\sin i \cos 2\pi\phi\rm{.}
  \end{equation}
%
%
%
  Finally, the flux of the reflected light from the planet at the orbital
  phase $\phi$ follows from Equations
  \ref{equ:2} and \ref{equ:3}:
  \begin{equation} \label{equ:4}
    f(\phi, i, \lambda) = \epsilon(\lambda)~\mu(\phi, i) \rm{.}
  \end{equation}
%
%

 \subsection{Doppler shifts} \label{dop2}


 The planet orbiting its host star does not only produce a flux variation 
 (Equation \ref{equ:4}), but also a Doppler shift of the
 stellar spectrum reflected from the planet. The RV semi-amplitude $K_{\rm p}$ of that shift
 depends on the orbital inclination $i$, which is unknown for non-transiting
 planets. $K_{\rm{p}}$ can be expressed by
  \begin{equation} \label{equ:doppler}
    K_{\rm{p}} = K_{\rm{s}} \frac{M_{\rm{s}}}{M_{\rm{p}}~\sin i} \sin i
    \rm{,}
  \end{equation}
  where $K_{\rm{s}}$ is the RV semi-amplitude of the star, and
  $M_{\rm{s}}$ and $M_{\rm{p}} \sin i$ are the stellar mass and the minimum
  mass of the planet,
  respectively. The largest possible amplitude $K_{\rm{p,max}}$ occurs at
  orbital inclination $i=90^{\circ}$ (Figure~\ref{F6:rv}).

  The instantaneous RV shift of the planetary signal with respect to the star depends 
  on the orbital phase
  $\phi$, 
  \begin{equation} \label{equ:doppler1}
    V_{\rm{p}} = K_{\rm{p}} \sin 2\pi\phi
    \rm{.}
  \end{equation}

\subsection{Rotational broadening} \label{2:rotbro}

As the star rotates, each absorption line is subjected to
Doppler broadening since individual points on the visible disk of the star have different 
instantaneous radial velocity. Equation~\ref{E2:rot1} describes the
projected rotational velocity of the star:
 \begin{equation} \label{E2:rot1}
v_{\rm proj,\star}=v_\star \sin i = 2 \pi \sin i~ \frac{ R_{\star}}{P_{\rm rot,\star}} ~,
 \end{equation}
where $R_{\star}$ and $P_{\rm rot,\star}$ are the stellar radius and the
rotation period of the star, respectively.
 If the inclination of the rotational axis is unknown, the true velocity $v$ of the star remains unknown.
 
Mathematically, the rotational broadening constitutes a convolution of the intrinsic
stellar line profile with a half ellipse whose width is equal to $\pm v~\sin i$ 
(Gray 1992). 
 The $v~\sin i$ of the reflected stellar absorption lines, as they could be
observed from the 
planet, can be calculated by way of Equation~\ref{E2:rot}:

 \begin{equation} \label{E2:rot}
v_{\rm \star,p}=2 \pi R_{\star} \left( \frac{1}{P_{\rm rot,\star}} -
  \frac{1}{P_{\rm orb}} \right) ~,
  \end{equation}
where $P_{\rm orb}$ is the orbital period of the planet. Equation~\ref{E2:rot} is based on the assumption that the orbital
 plane and the stellar equator are co-aligned. Analyses of Rossiter-McLaughlin effect RV-curves (Rossiter 1924;
McLaughlin 1924) in transiting planets have shown that such a spin-orbit alignment is the case for the majority of the monitored systems
(Fabricky \& Winn 2009), although such systems have been recently discovered (e.g. Pont et al.~2009, Narita et al.~2009).


To calculate the broadening coming from the planetary rotation, we use Equation~\ref{E2:rot111} 
\begin{equation} \label{E2:rot111}
v_{\rm proj, \rm p}=v_{\rm p} \sin i = 2 \pi \sin i~ \frac{ R_{\rm p}}{P_{\rm rot, p}} ~,
\end{equation}
where $R_{\rm p}$ is the planetary radius and $P_{\rm rot, p}$ the planetary rotation period. We note that this equation is 
based on the assumptions that the planetary rotation axis is co-aligned with the orbital inclination axis, which is furthermore co-aligned
with the stellar rotational axis. We justify these assumptions by
the following argument: Close-in planets such as hot Jupiter seem to be tidally locked with their parent stars 
(c.f.~Shkolnik et al. 2005; Knutson et al.~2007). A tidal lock hypothesis suggest that the orbital inclination is co-aligned with the
planetary rotation axis and with the stellar rotation axis. The latter assumption was confirmed by the aforementioned spin-orbit measurements of
transiting planets.

The two contributions $v_{\rm \star,p}$ (Equ.~\ref{E2:rot}) and $v_{\rm proj, \rm p}$ (Equ.~\ref{E2:rot111}) to the broadening are finally summed up to
 \begin{equation} \label{E2:rotz}
v_{\rm refl}=\sqrt{v_{\rm \star,p}^2+v_{\rm  proj, \rm p}^2}~.
 \end{equation}


\section{$\tau$ Boo and its planet} \label{S:6:2}
Table~\ref{tab:tauboo} summarizes the
  parameters of the planet and its host star. 
  We note in passing that the system also contains a faint M-dwarf
  component at a separatation of $\approx 230~\rm{AU}$ from the primary
  (Patience et al.~2002; Eggenberger et al.~2003).

\begin{figure}[t!]
   \centering
   \includegraphics[angle=-90,width=9cm]{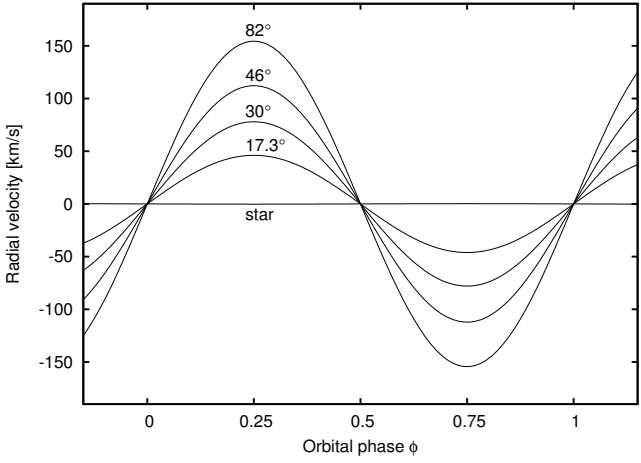}
      \caption{  RV curves of the planet orbiting $\tau$~Boo for selected
   orbital inclinations $i$. The maximum
   possible RV-shift of $K_{\rm p}=156~{\rm km~s^{-1}}$ occurs at $i=90^{\circ}$. For
   lower inclinations, the mass of the companion has to be larger  to
   produce the stellar reflex motion with an RV semi-amplitude
   $K_{\rm s}=0.4611~{\rm km~s^{-1}}$. { The most probable orbital inclination is $i=46^{\circ}$ under the assumption of a
   tidally locked system. The orbital inclination of $i=82^{\circ}$ is the upper limit, coming from the non-transit detection,
   whereas $i=30^{\circ}$ is the lower limit.}  For
   $i\le17.2^{\circ}$, the companion would exceed the mass of $13~~M_{\rm Jup}$ and therefore
   be a brown dwarf. }
          \label{F6:rv}
   \end{figure}
   
  \begin{table}
    \caption{Parameters of the star $\tau$~Boo and its planetary
  companion. Abbreviations for the references are: B06 = Butler et al. (2006) and references therein,
     B97~=~Baliunas et al. (1997), 
     G98~=~Gonzalez (1998),
     H00~=~Henry et al. (2000),
     L89~=~Latham et al (1989),
     L03~=~Leigh et al. (2003a) ,
     VF05~=~Valenty \& Fischer (2005).}             
    \label{tab:tauboo}      
    \begin{center}                         
    \begin{tabular}{l r l l }        
      \hline\hline                 
      Parameter & Value & Error &Ref. \\
      \hline                        
      Star: \\
      Spectral type & F7 IV-V  & & B97 \\ 
      $V~(mag)$        & 6.35 & & VF05\\
      $d~ (\rm{pc})$ & 15.60 & 0.17 & VF05\\
      $M_{\star}~ (\rm{M_{\odot}})$ & 1.33 & 0.11 & VF05 \\
      $R_{\star}~(\rm{R_{\odot}})$ & 1.31 & 0.06 & G98 \\
      $P_{\rm{rot}}~({\rm d})$ & 3.31 &  & B97 \\
      $v~ \sin i~ (\rm{km~s^{-1}})$ & 14.9 & 0.5 & H00 \\
      Age (Gyr) & $1.3$  & 0.4 & VF05 \\
      \hline
      Planet:\\
      $M_{\rm{p}} \sin i~(\rm{M_{\rm{Jup}}})$  & 4.13 & 0.34 & B06 \\
      $a~ (\rm{AU})$ & 0.0481 & 0.0028 & B06 \\
      $e$ & 0.023 & 0.015 & B06 \\
      $K_{\star}~(\rm{km~s^{-1}})$ & 0.4611 & 0.0076 & B06 \\
      Orbital period (d) & 3.31245 & 0.00003  & L03 \\
      ${ T_{\phi=0}}$ (JD) & 2~451~653.968 & 0.015  & L03 \\
      \hline                                   
    \end{tabular}
    \end{center}
  \end{table}
%

 \subsection{Radial velocity amplitude and orbital inclination}
  Knowing the stellar mass $M_{\star}$, the planetary minimum
  mass $M_{\rm{p}}\sin i$, and the
  RV semi-amplitude of the reflex motion of the star $K_{\star}$, 
  the RV semi-amplitudes of the planet can be constrained to ranging from
  $K_{\rm{p}} = 0$ to $155.6\pm18.3~\rm{km~s^{-1}}$ (Equation~\ref{equ:doppler}; Figure~\ref{F6:rv}).

Due to the absence of transits
  in high-precision photometry (Henry et al. 2000), we can constrain the range of possible orbital inclinations.
  The minimum orbital inclination $i$ that a transit event occurs can be calculated by 
 \begin{equation} \label{E2:iii}
\cos i = \frac{R_\star+R_{\rm p}}{a}~,
 \end{equation}
 where $R_\star$ and $R_{\rm p}$ is the stellar and the planetary radius, respectively, and $a$ the semi-major axis. 
 Using the parameters listed in Table~1, and the assumption that $R_{\rm p}=1.2~{\rm R_{\rm Jup}}$, 
  we can exclude
  orbital inclinations of $i>82^{\circ}$ for $\tau$~Boo~b.

Baliunas et al. (1997) found that the star $\tau$~Boo rotates rapidly with a period
 commensurate with the orbital period of the
planet, suggesting tidal locking. This hypothesis seems to be very likely for close-in planets (e.g. Shkolnik et al. 2005; Knutson et al.
2007), but has not been observationally confirmed so far.  A tidal lock enables us to place an estimate on the 
orbital inclination (Equation~\ref{E6:rot1}) under the assumption that the stellar
equator and the orbital plane are co-aligned.  
 
In order to calculate the orbital inclination, we need to solve 
\begin{equation} \label{E6:rot1}
  \sin i=\frac{v_{\rm proj,\star}}{v_\star} = \frac{v_{\rm proj,\star} P_{\rm rot,\star}}{2 \pi R_{\star}} ~,
 \end{equation}
 where $v_\star$ is the measured rotational velocity of $\tau$~Boo ($v_{\rm proj,\star} = v_\star \sin
i = 14.9~{\rm km~s^{-1}}$) for the unknown orbital inclination $i$, and $v$ the true, but
unknown rotational velocity. This rotational velocity $v_\star$ can be estimated by way of
Equation~\ref{E2:rot1}, which requires the knowledge of the stellar radius and stellar rotation period.
Adopting the parameters listed in Table~\ref{tab:tauboo},
we determine the maximum rotational velocity of the star to be $v_\star=20.0\pm 10~{\rm km~s^{-1}}$.
This gives an orbital inclination of $i=46^{+36}_{-16}~^{\circ}$. The resulting
mass of the companion to $\tau$~Boo of  ${M_{\rm p}=5.7^{+2.6}_{-1.5}~~M_{\rm Jup}}$ would clearly
confirm the planet hypothesis.We note that the error values of $i$ and $M_{\rm p}$ correspond to the maximum error
range.

\subsection{Rotational broadening} \label{rotbro}


We investigated how the reflected spectrum from the planet would be affected by a tidal lock and planetary
rotation, using the following asumptions: The hot Jupiter orbiting $\tau$~Boo rotates in $\approx3.1$
days (1:1 resonance with the orbital motion), and the rotation axis is co-aligned with that of the star. 
In addition,
the rotation of the planet is assumed to be prograde, and the planetary radius
is $R_{\rm p}
= 1.2~R_{\rm Jup}$. 

A tidal lock causes that an imaginary observer sitting on the planets always sees the same side of the star; 
the star appears to be non-rotating.
For this reason, $v_{\rm \star,p}=0~{\rm km~s^{-1}}$. According to Equation~\ref{E2:rot111}, 
we find that the reflected
absorption lines from the planet are broadened by approximately
$v_{\rm p}~\sin i=2~{\rm km~s^{-1}}$. In order to estimate the resulting full width at half maximum (FWHM) of the reflected
stellar absorption lines, we investigate slowly rotating stars of similar
spectral type. 

A very good candidate is the star HD~136351  (spectral type: F6 IV), which shows a
rotational broadening of $v_{\star} \sin
i=2.8~{\rm km~s^{-1}}$ (Reiners \& Schmidt 2003). The average FWHM of the absorption lines
is $12~{\rm km~s^{-1}}$. Since the rotational velocity of the planet is estimated to be smaller
(see above), we estimate the resulting FWHM of the reflected stellar
absorption lines to be about $11.8~{\rm km~s^{-1}}$.

\section{Data acquisition and data reduction} \label{S:6:5}
 
We observed $\tau$~Boo during two consecutive nights in June 2007 using UVES
(Dekker et al.~2000)  mounted
   on the VLT/UT2 and collected a total of 
406 high-reolution spectra (Table~\ref{t6:journal}). In addition to that, we 
took spectra of the slowly rotating star HD~136351 and the rapidly  rotating B-star HD~116087 
($v_\star \sin i=241~{\rm km~s^{-1}}$; Hoffleit \& Jaschek 1991).
The latter star was observed for the identification of telluric features in the
red part of the visual. The dates of the observations were selected in such a way that the observations were carried out at 
orbital phases at which the planetary signal was
strongest, i.e. close to the position of superior conjunction, which would be the secondary transit in
case of $i=90^\circ$. We aimed at taking observations in the phase ranges $\phi=0.30$ to $0.45$ and $0.55$ to
$0.70$, but avoiding $\phi=0.50$ because of intense blending of the planetary and stellar absorption lines.

The observations were conducted using the red arm of UVES with a non-standard setting 
centered at wavelength $\lambda = 430~\rm{nm}$ providing 47 full
   orders and covering the wavelength range $\lambda=425$ to $632~\rm{nm}$. 
We observed through a 0.3'' slit and the image slicer IS\#3, providing us a
spectral resolution of $R=\lambda/\Delta\lambda=110~000$.
The integration time for our target was selected to provide the
maximum count rate but  at the same time avoiding to reach the 
non-linear regime of neither one of the two CCDs. The
exposure times for our bright target ranged between 60 and 400~s (120~s
on average) and 30 and 60~s (40~s on average), respectively for the first and the second night.  
We carried out our observations in the fast-R/O mode; the
total dead-time between two exposures was 16~s.
Due to clouds and thick cirrus, we only collected half of the expected spectra.

   \begin{table*}
     \caption{Journal of observations. The UTC times and the orbital phases of
   the planet are shown. Ephemerides used are:  
   BJD~$2451653.968 + 3.31245\times {\mathrm E}$ (Leigh et al.~2003a quoting Marcy priv.~comm.).}             

 \centering                          

     \begin{tabular}{c c c c c c c c  }        
       \hline\hline                 
       Night & UTC start &  BJD start & $\phi$ & UTC end &  BJD end & $\phi$ &  $N_{\rm{spectra}}$
       \\
       \hline                        
        1 & 2007/06/16~~23:24 & 2454268.4773 & 0.30  & 2007/06/17~~04:12 & 2454268.6784 & 0.36 &
        100\\
        2 & 2007/06/17~~22:45 & 2454269.4498 & 0.59  & 2007/06/18~~04:26 & 2454269.6873 & 0.66 &
        306\\
       \hline                                   
     \end{tabular}
     \label{t6:journal}      
  \end{table*}

In addition to the science frames, we obtained a large number of calibration
exposures. Before the start of the first night, we took 163 flat-field 
and 104 bias exposures. In the afternoon before the second night, we
collected 245 flat-field and 65 bias exposures. The large number of calibration frames was important to
   achieve a low photon noise in the finally combined flat-field frame (masterflat).

\subsection{Data reduction} 

In order to prevent introducing data reduction artefacts to the data, we aimed at
keeping the spectra as raw as possible. Consequently, we only adopted the
absolutely necessary data reduction steps, as pointed out in the following. Effects like errors in the
wavelength calibration, trends in the continuum, instrumental
profile changes were considered in the model for the star/planet (see Section~5).

In the first step, we created masterbias frames for each night, being the median of all bias
exposures per night. These masterbias frames were then subtracted  from the science frames as
well as from the flat-field frames. The error of the flux in the science frames
   was
   determined on the basis of Poisson statistics and
   propagated in the further data analysis steps. In the following step,  for each night the flat-field
   exposures were scaled with their inverse exposure times and then combined into their median (masterflat) to permit high-quality
   flat-field correction in order to compensate for sensitivity variations of
   the pixels.  
   The science frames were divided by the appropriate masterflat.
   Since both
   the flat-field frames as well as the object frames show a similar Blaze
   function, this function was then mostly removed from the object frame.
    Finally,
   1-dimensional spectral orders (pixel vs. flux) were extracted from the
   2-dimensional frames. 
 
  We retrieved 47 orders
   of 4096 pixels each. 
   No order merging was applied.  The observation of the UVES
   Thorium-Argon (Th-Ar) lamp enabled us to assign each pixel the appropriate value of the
   wavelength. For each night, we used the 
   Th-Ar spectrum observed before/after the science exposures and established an 8th order
   polynomial dispersion solution. 
   All these data reduction steps were performed with the MIDAS software package.

\section{Data analysis} \label{S:6:5a}

   After the data reduction, we identified cosmic-ray hits by way
   of the following procedure: for each spectrum, we
   compared the flux in every pixel with the median flux of the same pixel in
   the three predecessor and the
   three successor spectra, which had been scaled to the same flux as the
   spectrum under consideration. We flagged those pixels where the 
   difference exceeded $6 \sigma$ as cosmic-ray hits. These pixels were 
   then excluded from further analysis.   We furthermore discarded the most weakly exposed regions of each echelle order 
   (the first 400 pixels as well as the last 240
   pixels) since the flux level of the continuum was subjected to strong
   variations. In addition,  the spectral  order around the wavelength of
   $\lambda=630$~nm was excluded from further analysis, since data were affected by the strong telluric OI features.
   In order to speed up the data analysis, we co-added spectra into
   sufficiently narrow phase bins in such a way that phase smearing of the stellar absorption lines 
originating from stellar RV
   variation, sub-pixel shifts and the
   effect of barycentric motion of the Earth remained negligible. The main criterion for the size of the phase bins was that the
   unseen planetary lines did not suffer from smearing in excess of $2~{\rm km~s^{-1}}$.
   With that step,
   we were able to reduce the number of spectra from 406 to 58.

\subsection{Data synthesis method} \label{S:6:5b}
   
   We model the starlight reflected from the planet as a copy of the stellar spectrum,
   strongly scaled down in brightness and Doppler-shifted according to the orbital motion of the
   planet. The target spectra have an average S/N of $300$ to $600$ per
   spectral bin. With expected planet-to-star flux ratios
   of the order of a few times $10^{-5}$, it is clear that the reflected spectrum from 
   the planet is deeply buried in the noise of the stellar spectrum. The weak planetary
   signal is boosted by the large number of spectra, and more importantly, 
   by the combination of the approximately $1500$ absorption lines, achieved using the 
   data-synthesis method (cf. Charbonneau et al.~1999) described below in a nutshell (for a detailed description see Rodler et al. 2008).

\begin{figure}[t!]
   \centering
   \includegraphics[angle=-90,width=9cm]{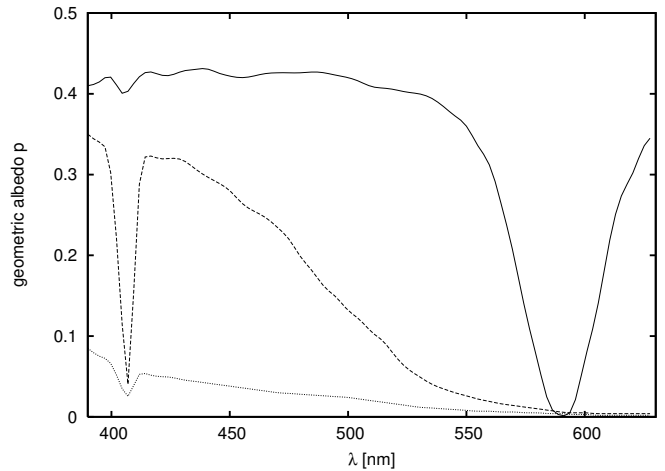}
      \caption{Different albedo spectra of atmospheric models (taken from Sudarsky et
   al. 2000) are shown. The irradiated (dots) and isolated (dashed) Class IV
               models describe atmospheres of planets with temperatures 
               $T_{\rm{eff}}\approx1300~\rm{K}$. Contrary to the isolated model,
               the irradiated model
               assumes that no reflective clouds exist in the upper layer of
               the planetary atmosphere, which results in a very low
               albedo. 
               The solid line depicts a Class V roaster, describing
               the atmosphere of a planet with $T_{\rm{eff}}\ge1500~\rm{K}$, having a high
               reflective silicate cloud deck in the upper layers of the atmosphere. This atmospheric model was already ruled out for $\tau$~Boo~b by
			   Leigh et al.~(2003a).}
          \label{sudarsk}
   \end{figure}

In the first step, a high S/N, virtually planet-free superspectrum is computed by co-adding the observed spectra 
after correcting
their wavelength values for the barycentric motion of the Earth and for the stellar orbital motion.
 Then, we construct a model to describe each of the original, unmodified object spectra: 
the dominant stellar signal is represented by 
the superspectrum, shifted to the position of the observed spectrum. Imperfections in flux and wavelength are corrected by using the approach described in 
Rodler et al.~(2008). Furthermore, the contribution to the spectrum of the planetary signal is created as follows: we   
 adopt the spectrum of the slowly rotating F6~IV
star HD~136351 to mimic the expected sharp reflected lines of the planet (Section~\ref{rotbro}). 
For each observed object spectrum, we need to co-align that spectrum of
HD~136351 with the stellar model spectrum (shifted superspectrum).
  The co-aligned spectrum of HD~136351 is then scaled down by the factor  $\epsilon(\lambda)~\mu(\phi, i)$ and
     shifted by velocity $V_{\rm{p}}(K_{\rm{p}},\phi)$ with respect to the
     stellar spectrum. Hence, the two free parameters of the planetary model are the planet-to-star
     flux ratio for the fully-illuminated planet $\epsilon(\lambda)$, and the
     orbital inclination $i$, which corresponds to an 
     RV semi-amplitude of the planet $K_{\rm{p}}=K_{\rm{p,max}} \sin i=155.6
     \sin i~~{\rm km~s^{-1}}$.  Concerning the albedo function $p(\lambda)$, we adopt (i) a grey albedo 
  assumption (i.e. $p(\lambda)={\rm constant}$ for all wavelengths $\lambda$) and (ii) the irradiated 
  atmospheric Class IV model by Sudarsky
   et al.~(2000), which predicts higher reflectivity at shorter wavelengths (Fig.~\ref{sudarsk}). 

  The search range for the RV
  semi-amplitude comprises $K_{\rm p} = 40$ to $180~{\rm km~s^{-1}}$
  (corresponding to orbital inclinations $i=15^\circ$ to $90^\circ$, plus
  twice the error of $K_{\rm p,max}$ (i.e. $18.3~{\rm km~s^{-1}}$ with a step
  width of $3~{\rm km~s^{-1}}$. This is a good compromise between computing
  time and sampling the  expected average reflected absorption line profile with the FWHM of $\approx
  12~{\rm km~s^{-1}}$. Since the tidal lock hypothesis and a spin-orbit
  alignment is very likely,
  low-inclination orbits can be furthermore excluded: the observed $v \sin i = 15~{\rm km~s^{-1}}$ would imply
an improbable true rotational velocity above $50~{\rm km~s^{-1}}$ for an F7~IV-V star at very low orbital inclinations
(c.f. rotational velocities of late F-type stars; Glebocki \& Stawikowski~2000).
Using simulations, we found that for small
  inclinations of the planetary orbit, where the planets appear only slightly
  illuminated, and the method would fail to detect Jupiter-size objects even with very high albedos.
 
   For the second parameter $\epsilon(\lambda_0)$, we
 search for the planetary signal in the range  $10^{-4}$ to $10^{-5}$ with a
 step size of $5\times10^{-6}$. 

   \subsection{Determination of the confidence level}
   Once the best model $M[K_{\rm p},\epsilon(\lambda)]$ has been evaluated, 
   we determine the confidence level of the $\chi^2$ minimum. 
   We do not infer the confidence from the probability of the minimum $\chi^2$ value, because 
   due to unknown systematic errors coming from the telescope
  and instrument the true errors are largely underestimated when
  considering only photon statistics and detector read-out noise.
      Therefore no reasonable probability estimate is possible
      from the $\chi^2$ value alone. 
   For this reason, we
   apply the
   bootstrap randomization method (e.g. K\"urster et al.~1997). Retaining the
   orbital phases, we randomly redistribute the observed
   spectra amongst the phases, thereby creating a large number ($N=3000$) of different data sets.
   Any signal present in the original data is now scrambled in these
   artificial data sets.
   For all these randomised data sets, we again evaluate the model for the
   two free parameters, and locate the best fit with its specific $\chi^2$
   minimum. We set $m$ to be the number of best-fit models to the $N$ randomised
   data sets that have a minimum $\chi^2$ less or equal than the minimum $\chi^2$ found for the original
   data set. The confidence level can then be estimated by $1-m/N$.

\section{Results} \label{sec:6:6}
\subsection{Grey albedo model} \label{6:grey}
 \begin{figure}
  \centering \includegraphics[bb=30 50 550
  780,angle=-90,width=9cm]{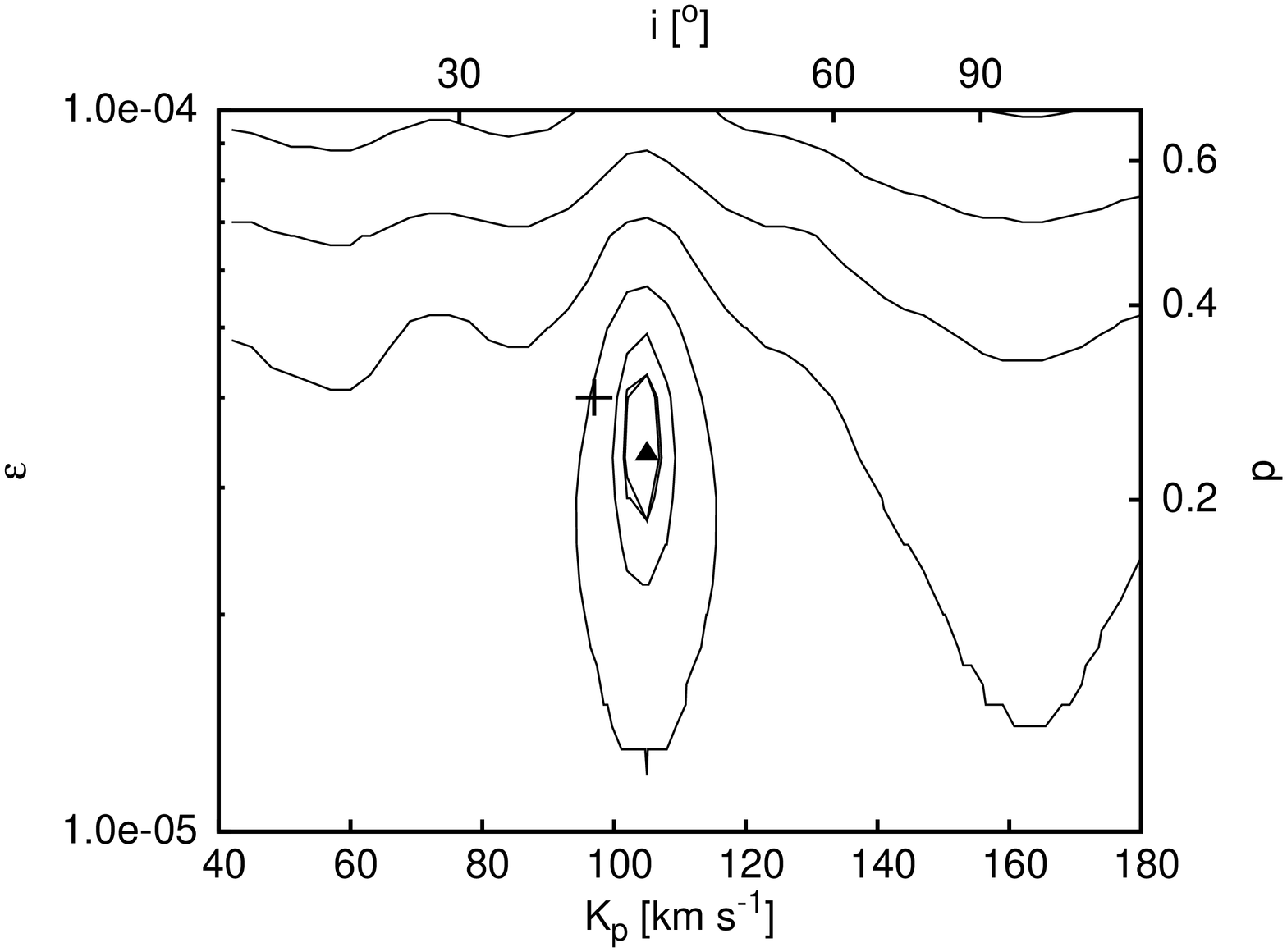}
  \includegraphics[bb=30 50 550 780,angle=-90,width=9cm]{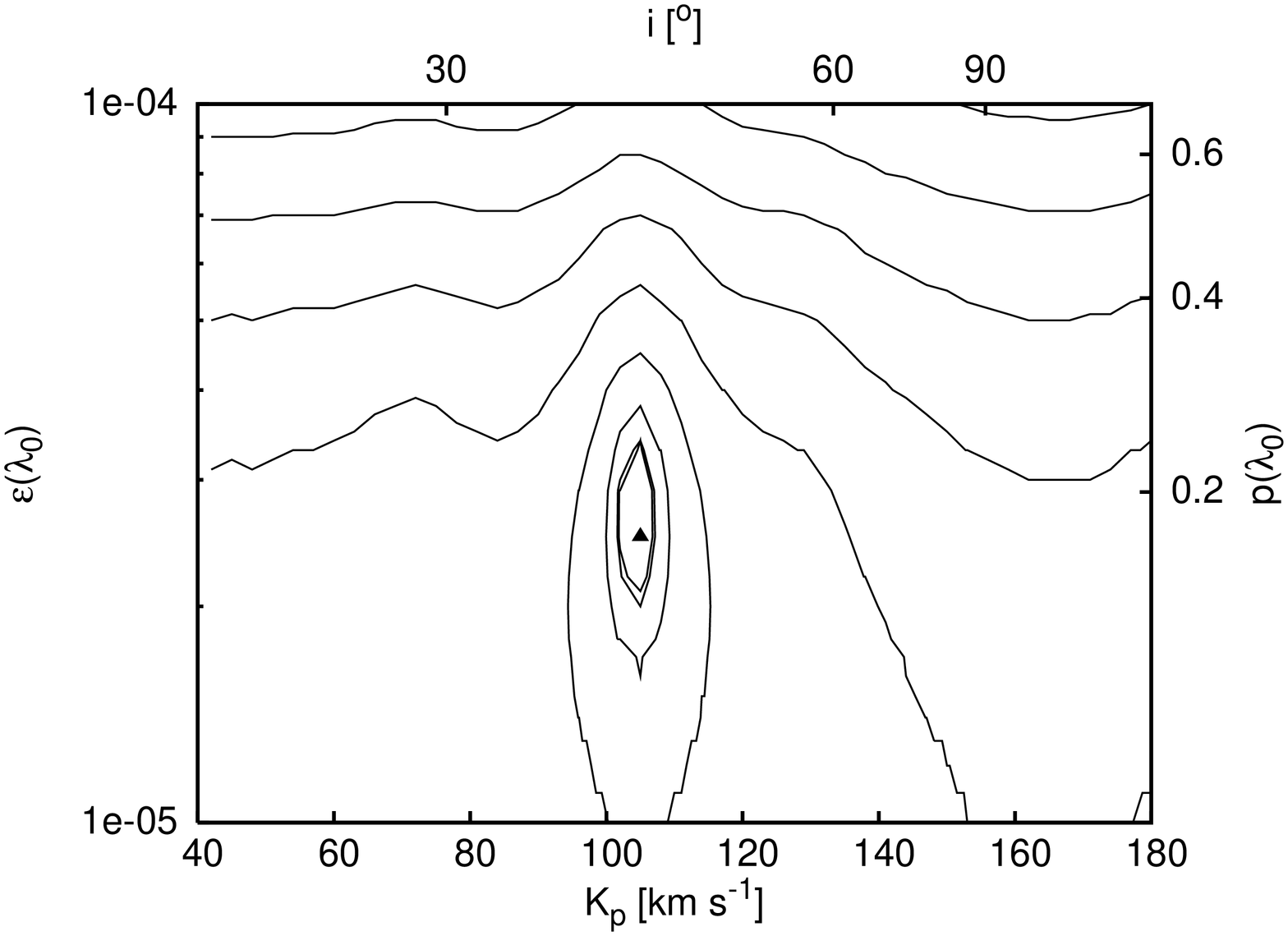}
      \caption{Contour maps of $\chi^2$ for 
        the model parameters of the planet of $K_{\rm{p}}$ (RV semi-amplitude of the planet) and
   $\epsilon(\lambda_0)$ (planet-to-star flux ratio); this plot shows the result of the analysis of the $\tau$~Boo~b data for two different atmospheric models.
  The contour levels follow the
  sequence $C_k=\chi^2_{\rm
     min}+10^{-6} (\chi^2-\chi^2_{\rm
     min})^3$ for $k>0$, where $\chi^2_{\rm min}$ is the minimum $\chi^2$.  \newline
        Upper panel: grey albedo model (model \#1). Lower panel: irradiated Class IV albedo function (model \#2).
   For both cases, a Venus-like phase function was adopted.    
   The minimum $\chi^2$ was found at
  $K_{\rm{p}}=103~\rm{km~s^{-1}}$ and
  $\epsilon=3.3\times10^{-5}$, and $K_{\rm{p}}=103~\rm{km~s^{-1}}$ and
  $\epsilon(\lambda_0)=2.5\times10^{-5}$, respectively, for model \#1 and model \#2.  
  A bootstrap analysis with 3000 trials revealed that these features are uncertain with
  a false alarm probability of $3.6\%$ and $7.3 \%$, respectively, for model \#1 and model \#2.
   For comparison, we show the candidate feature found by Leigh et al.~(2003a) at $K{\rm{p}}=97~\rm{km~s^{-1}}$ 
  and $\epsilon=4\times10^{-5}$, assuming the grey albedo case (upper panel, cross). 
  This candidate feature has a FAP of $1.4 \%$.
 Note that the corresponding geometric albedo $p$ is shown under the
   assumption that $R_{\rm p} = 1.2~R_{\rm Jup}$ (the right-hand
   y-axis).
 }
         \label{fig:nodetection}
   \end{figure}


   Our data analysis adopting a grey albedo model resulted in  
   a $\chi^2$-minimum at a planet-to-star flux ratio $\epsilon =
   3.3\times10^{-5}$ and an RV semi-amplitude
   $K_{\rm{p}}=103~\rm{km~s^{-1}}$ corresponding to an orbital inclination of
   $i=41^{\circ}$. Figure \ref{fig:nodetection} (upper panel) shows the 
 $\chi^2$ contour map in the $\epsilon$ vs. $K_{\rm p}$ plane. However, using bootstrap randomisation with 3000
   trial data sets we find  this $\chi^2$-minimum to be uncertain with a
   false alarm probability (FAP) of 3.6\%, and consequently do not consider it
   as a detection of reflected light from the planet.

 \subsection{Irradiated Class IV model function}

   The analysis with the irradiated Class IV albedo model 
   did not yield  any evidence  for reflected light either.
    We find the formal $\chi^2$-minimum at a planet-to-star flux 
   ratio $\epsilon(\lambda_0) = 2.5 \times
   10^{-5}$ and an RV semi-amplitude of
   $K_{\rm{p}}=103~\rm{km~s^{-1}}$, which again corresponds to an orbital
   inclination $i=41^\circ$. Figure \ref{fig:nodetection} (lower panel) shows the 
 $\chi^2$ contour map in the $\epsilon$ vs. $K_{\rm p}$ plane. From bootstrap randomisation with 3000
   trials we found this feature to be uncertain with a FAP of 7.3\%. 
   

\subsection{Upper limits}
  \begin{figure}[t!]
   \centering \includegraphics[bb=30 50 550 780,angle=-90,width=9cm]{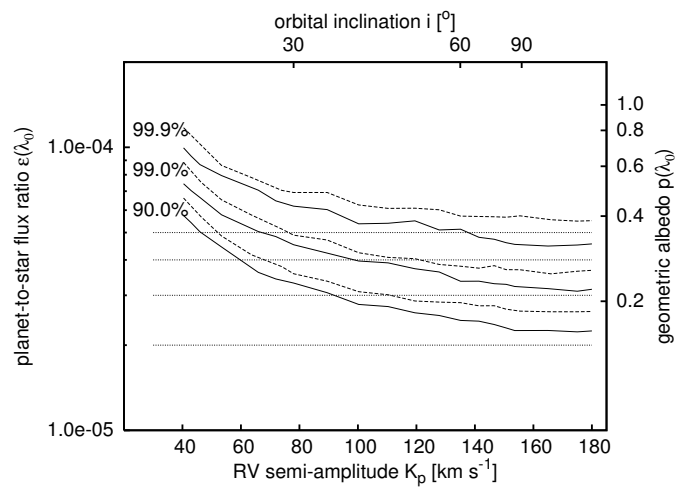}
   \vspace{5mm}
   \caption{Contour map showing confidence levels for the upper limits to
   the planet-to-star flux ratio $\epsilon(\lambda_0)$ as a function of to the 
   RV semi-amplitude
   $K_{\rm{p}}$ of the planet (lower x-axis), or orbital inclination $i$
   (upper x-axis). 
   From top to bottom, these levels are:
   99.9, 99.0 and 90.0 \% confidence. For the planetary model we assumed a Venus-like
   phase function as well as a irradiated Class IV albedo function (dashed lines) and a grey
   albedo (solid lines). The corresponding geometric albedo $p$ is shown under the
   assumption that $R_{\rm p} = 1.2~R_{\rm Jup}$ (the right-hand
   y-axis). 
}
         \label{f6:upp}
   \end{figure}

  Given the non-detection of the planetary signal, we determined upper limits of the planet-to-star flux
   ratio $\epsilon(\lambda_0)$ for different confidence levels as a function of 
   the RV semi-amplitude $K_{\rm p}$. Figure~\ref{f6:upp} shows that
   the upper limits to the planet-to-star flux
   ratio decrease with increasing orbital inclination, which is a direct
   consequence of the illumination geometry. 
   As can be seen from Figure~\ref{f6:upp} (solid lines), we have
   the highest sensitivity for detection of the planetary signal at high
   orbital inclinations. For the most probable orbital inclination of $46^{\circ}$, the 99.9~\% confidence
   upper limit to the planet-to-star flux ratio for the grey albedo model is
   $\epsilon=5.7\times10^{-5}$ , while it is  $\epsilon=6.5\times10^{-5}$  for
   the irradiated Class IV albedo function. For an orbital inclination of $i=60^{\circ}$, 
the 99.9~\% confidence
   upper limit to the planet-to-star flux ratio for the grey albedo model is already
   $\epsilon=5.1\times10^{-5}$ , while it is  $\epsilon=5.7\times10^{-5}$  when
   adopting the irradiated Class IV albedo function.
Similar to our results for the hot Jupiter HD~75289Ab (Rodler et al.~2008), Figure~\ref{f6:upp} shows 
   that the upper limits established by adopting the grey albedo model are
   deeper than the ones found with the irradiated Class IV model. 

   Assuming a planetary radius $R_{\rm p}=1.2~R_{\rm Jup}$ and an orbital
   inclination $i\approx46^{\circ}$ (cf. Section~\ref{S:6:2}), we find the
   upper limit to the geometric albedo to be $p<0.40$ for the grey albedo
   model and  $p<0.44$ for the irradiated Class IV model at a wavelength 
   of $\lambda_0=462~{\rm nm}$, which corresponds to the centre of gravity of the irradiated
   Class IV albedo function in the observed wavelength range. 
   For comparison, at this wavelength Jupiter's
   geometric albedo is $p=0.44$
   (Karkoschka 1994). 


\section{Discussion and conclusions} \label{sec:6:7}

We have observed the hot Jupiter orbiting $\tau$~Boo for two half nights
with UVES, mounted at the VLT/UT2, in an attempt to measure starlight
reflected from the planetary companion. 
\begin{itemize}
   \item The data analysis using the data synthesis method did not reveal a
   detection of the reflected light from the planet. The signal-to-noise ratio of the combined data 
   was not sufficient to significantly detect the planetary signal; 
instead we placed
   upper limits to the planet-to-star flux ratio for
   different possible orbital inclinations and confidence levels. {Under the assumption of a planetary radius of
   $R_{\rm p}=1.2~R_{\rm Jup}$, we rule out atmospheric models with geometrical albedos $p>0.40$ for $\tau$~Boo~b. }
   We confirm the finding by Leigh et al. (2003a) that the  
   upper limit to the geometric albedo of the planet (e.g. $p<0.39$ for orbital
   inclinations $i \geq 36^{\circ}$ and grey albedo) implies that this
   planet has a lower reflectivity than the atmospheres of Jupiter ($p=0.44$)
   and Saturn ($p=0.62$).\\

  \item For both albedo models, we find a formal $\chi^2$-minimum at $K_{\rm{p}}=103~\rm{km~s^{-1}}$.
For the analysis adopting the irradiated Class IV albedo model, we determine the FAP of the candidate feature to be 7.3\%. In comparison, Leigh et al. (2003a) finds
a candidate feature of marginal significance (FAP=0.9\%) with $K_{\rm
  p}=90~{\rm km~s^{-1}}$.

For the analysis adopting the grey albedo model, we find that this candidate feature has a low FAP of
3.6\% with a planet-to-star flux ratio of $\epsilon=3.3\times10^{-5}$.
If genuine, this would correspond to an
orbital inclination of $i=41^{\circ}$ and a geometric albedo of $p=0.23$
assuming  $R_{\rm p}=1.2~R_{\rm Jup}$. For the grey albedo model, Leigh et al.~(2003a)
reports about the detection of a non-significant candidate feature with
a similar value of $K_{\rm
  p}=97~{\rm km~s^{-1}}$, having a low FAP of $1.4\%$.

These similar, but non-significant candidate features were found by
using completely independent
data sets and data analysis methods. Leigh et al.~(2003a) analyzed high-resolution spectroscopic data, which had been 
obtained during 17 nights 
with the Utrecht Echelle Spectrograph (UES), mounted on the William Herschel-telescope in La Palma, Spain. 
Contrary to our analysis
strategy, they used a deconvolution
approach (Collier Cameron et al. 2002) for the extraction of the planetary signal. In this work, we analyzed UVES data, obtained during
two half nights, by using the data synthesis method. These results are also consistent with an estimate of the orbital incliation
of $i\approx46^\circ$ (corresponding to a semi-amplitude of $K_{\rm
  p}=112~{\rm km~s^{-1}}$) and assuming 
tidal locking.

{Leigh et al.~(2003a) achieves a higher sensitivity by adopting the irradiated Class IV model atmosphere, whereas our
highest sensitivity is attained with the grey albedo assumption. One explanation for this might be that Leigh et al.~(2003a) were using data 
with a slightly better wavelength coverage at blue
wavelengths (407-647 nm), while our data set covered the wavelength range from 425 to 632 nm. However, it seems to be more likely that this discrepancy
between the best-fit atmospheric models is just coincidence, since both campaigns resulted only in non-significant candidate features.
}


  \end{itemize}

  \begin{acknowledgements}
  We are very grateful to Tsevi Mazeh 
   for valuable discussions.

   \end{acknowledgements}



\begin{thebibliography}{}
\bibitem[2009]{al09a} Alonso R., Alapini A., Aigrain S., et. al., 2009a, A\&A, 506..353
\bibitem[2009]{al09b} Alonso R., Guillot T., Mazeh T., et. al., 2009b, A\&A, 501, 23

\bibitem[2007]{bakos07} Bakos G., Noyes R., Kov\'acz G., Latham D., et
  al. 2007, ApJ Letters, 656, 552

\bibitem[1997]{bal97} Baliunas S., Henry G. W., Donahue R. A., Fekel F. C., \& Soon W. H. 1997, ApJ, 474, L119

	
 \bibitem[1997]{but97} Butler R. P., Marcy G. W., Williams E., et al. 1997, ApJ, 474, L115

    \bibitem[2006]{butler06} Butler R. P., Wright J. T., Marcy G. W., Fischer
     D. A., Vogt S. S., Tinney, C. G., 2006, ApJ, 646, 505

   \bibitem[1999]{charb} Charbonneau D., Noyes R.W., Korzennik S.G., Nisenson
   P., Jha S., Vogt S.S., \&~Kibrik R.I., 1999, ApJ, 522, L145

\bibitem[1999]{col99} Cameron A.~C., Horne K.,  Penny A., James D. 1999, Nature, 378, 355
\bibitem[2009]{chr09} Christiansen J. L., Ballard S., Charbonneau D., et al., 2010, ApJ, 710, 97
\bibitem[2002]{col02} Collier Cameron A., Horne K., Penny A., Leigh C. 2002, MNRAS,330, 187
\bibitem[2010]{cow10} Cowan N. B. \& Agol E., 2010, arXiv:1001.0012
 \bibitem[2000]{dek00} Dekker H., D'Odorico S., Kaufer A., et al., 2000, SPIE, 4008, 534.

\bibitem[2003]{Egg03} Eggenberger A., Udry S., \& Mayor M. 2003, ASP Conf. Ser., 294, 43

\bibitem[2009]{Fab09} Fabricky \& Winn 2009, arXiv: 0902.0737

\bibitem[2000]{gle00} Glebocki R. \&  Stawikowski A., 2000, AcA, 50, 509

\bibitem[1998]{gon98} Gonzalez G. 1998, A\&A, 334, 221

 \bibitem[1989]{gra89} Gratton R.~G., Focardi P., Bandiera R., 1989, MNRAS, 237, 1085
 \bibitem[1992]{gra92} Gray, 1992, "The observation and analysis of stellar photospheres", Camb.~Astrophys.~Ser., Vol.~20
 Astropysics


   \bibitem[2000]{hen00} Henry G. W., Baliunas S. L., Donahue R.~A., et al., 2000, ApJ, 531, 415

   \bibitem[1992]{hilton92} Hilton, J.L., 1992, Explanatory Supplement to the
     Astronomical Almanac, Universiy Science Books, Mill Valley CA, p. 383
 
  \bibitem[1991]{hof91} Hoffleit D. \& Jaschek C., 1991, The Bright star catalogue,  Yale University Observatory, New
  Haven, Connecticut

   \bibitem[1994]{Karkoschka} Karkoschka E., 1994, Icarus, 111, 174

\bibitem[2007]{Knu07} Knutson H., Charbonneau D., Allen L. E., et al. 2007, Nature, 447, 183

   \bibitem[1997]{Kur97} K\"urster M., Schmitt J.H.M.M., Cutispoto G., Dennerl
   K., 1997, A\&A, 320, 831

\bibitem[1989]{lat89} Latham D. W., Stefanik R. P., Mazeh T. Mayor M. \& Burki G. 1989, Nature, 339, 38

   \bibitem[2003a]{leigh} Leigh C., Collier Cameron A., Horne K., Penny A., James D., 2003a, MNRAS, 344, 1271

   \bibitem[2003b]{leigh} Leigh C., Collier Cameron A., Udry S., Donati J.-F., Horne K., James D., Penny
   A. 2003b, MNRAS, 346, 16







\bibitem[1924]{McL24} McLaughlin D. B. 1924, ApJ, 60, 22
 \bibitem[2009]{Narita09} Narita N., Sato Bun\'ei H. T., Tamura M., 2009, PASJ, 61, L35

\bibitem[2002]{Pat02} Patience J., White R. J., Ghez A. M., et al. 2002, ApJ, 581, 654

\bibitem[2009]{Pon09} Pont F., H\'ebrard G., Irwin J. M., Bouchy F., 2009, A\&A, 502, 695

 
  \bibitem[2008]{Rei03} Reiners A. \& Schmidt  J. H. M. M. 2003, A\&A, 412, 813
 
 \bibitem[2008]{Rod08} Rodler F., K\"urster M., Henning T. 2008, A\&A, 485, 859

\bibitem[2009]{rog09} Rogers J. C., Apai D., L\'opez-Morales M., et al., 2009, ApJ, 707, 1707
 \bibitem[1924]{Ros24} Rossiter R. A. 1924, ApJ, 60, 15
 \bibitem[2008]{Row08} Rowe J. F., Matthews J. M., Seager S., et al., 2008, ApJ, 689, 1345
  
   \bibitem[1998]{sea98} Seager S. \& Sasselov D. D., 1998, ApJ, 502, 157

\bibitem[2005]{shk05} Shkolnik E., Walker G. A. H., Bohlender D. A., Gu P.-G., Kürster M., 2005, ApJ, 622, 1075
\bibitem[2009]{sne09} Snellen I. A. G., de Mooij E. J. W.. Albrecht, S., 2009, Nature, 459, 543

\bibitem[2009]{slm09} Sing D. K. \&  L\'opez-Morales M., 2009, A\&A, 493, 31

 \bibitem[2000]{Sudarsky00} Sudarsky D., Burrows, A., Pinto P., 2000, ApJ,
   538, 885
        
  \bibitem[2000]{Sudarsky03} Sudarsky D., Burrows A., Hubeny I., 2003, ApJ,
  588, 1121

   \bibitem[2005]{VF05} Valenti J. A., Fischer D. A., 2005, ApJ Supl., 159, 141
\end{thebibliography}
\end{document}